\documentclass[10pt]{article}
\usepackage{amssymb,latexsym,amsmath,amsthm,graphicx}
\usepackage{amscd}
\usepackage{epsf}

\makeatletter \@addtoreset{figure}{section}
\def\thefigure{\thesection.\@arabic\c@figure}
\def\fps@figure{h, t}
\@addtoreset{table}{bsection}
\def\thetable{\thesection.\@arabic\c@table}
\def\fps@table{h, t}
\@addtoreset{equation}{section}

\newif\ifamsfonts\amsfontstrue \ifamsfonts
\font\twlbbb=msbm10 scaled\magstep1 \font\egtbbb=msbm8
\font\sixbbb=msbm6
\newfam\mathbbfam
\textfont\mathbbfam=\twlbbb \scriptfont\mathbbfam=\egtbbb
\scriptscriptfont\mathbbfam=\sixbbb
\makeatother

\newtheorem{thm}{Theorem}[section]
\newtheorem{prop}[thm]{Proposition}

\newtheorem{remark}{Remark}[section]

\newcommand{\R}{\mathbb{ R}}
\newcommand{\h}{\mathfrak{ h}}
\newcommand{\g}{\mathfrak{ g}}
\newcommand{\D}{\mathfrak{ d}}
\newcommand{\N}{\mathcal{ N}}

\DeclareMathOperator{\pr}{pr}
\DeclareMathOperator{\Span}{span}
\DeclareMathOperator{\diag}{diag}
\DeclareMathOperator{\ad}{ad}
\DeclareMathOperator{\Ad}{Ad}
\DeclareMathOperator{\tr}{tr}

\begin{document}

\title{LR and L+R Systems}

\author{Bo\v zidar Jovanovi\'c}

\maketitle

\abstract{We consider coupled nonholonomic LR systems on the
product of Lie groups. As examples, we study $n$-dimensional
variants of the spherical support system and the rubber Chaplygin
sphere. For a special choice of the inertia operator, it is proved
that the rubber Chaplygin sphere, after reduction and a time
reparametrization becomes an integrable Hamiltonian system on the
$(n-1)$--dimensional sphere. Also, we showed that an arbitrary L+R
system introduced by Fedorov in \cite{Fe1} can be seen as a
reduced system of an appropriate coupled LR system.\footnote{AMS
Subject Classification 37J60, 37J35, 70H45} }

\baselineskip=12pt

\section{Introduction}

In this paper we study nonholonomic geodesic flows on direct
product of Lie groups with specially chosen right-invariant
constraints and left-invariant metrics.

Let $Q$ be a $n$--dimensional Riemannian manifold $Q$ with a
nondegenerate metric $\kappa(\cdot,\cdot)$ and let $\mathcal D$ be
a nonintegrable $(n-k)$--dimensional distribution on the tangent
bundle $TQ$. A smooth path $q(t)\in Q,\; t\in\Delta$ is called
{\it admissible} (or allowed by constraints) if  the velocity
$\dot q(t)$ belongs to ${\mathcal D}_{q(t)}$ for all $t\in\Delta$.
Let $q=(q_1,\dots,q_n)$ be some local coordinates on $Q$ in which
the constraints are written in the form
\begin{equation}
(\alpha^j_q,\dot q)=\sum_{i=1}^n \alpha_i^j\dot q_i=0,\qquad
j=1,\dots,k, \label{constraints}\end{equation} where $\alpha^j$
are independent 1-forms.
The admissible path $q(t)$ is called
{\it nonholonomic geodesic} if it is satisfies the Lagrange--d'Alambert equations
\begin{equation}
\frac{d}{dt}\frac{\partial L}{\partial \dot q_i}= \frac{\partial
L}{\partial q_i}+  \sum_{i=1}^{k}\lambda_j \alpha^j(q)_i,  \qquad
i=1,\dots,n,\label{Hamilton}
\end{equation}
where the Lagrange multipliers $\lambda_j$ are chosen such that
the solutions $q(t)$ satisfy constraints (\ref{constraints}) and
the Lagrangian is given by the kinetic energy $L=\frac12
\kappa(\dot q,\dot q)=\frac12\sum_{ij} \kappa_{ij} \dot q_i \dot
q_j$. After the Legendre transformation $p_i={\partial L}/\dot
q_i=\sum_j \kappa_{ij} \dot q_j$, $i=1,\dots,n$, one can also
write the Lagrange-d'Alambert equations as a first-order system on
the cotangent bundle $T^*Q$. As for the Hamiltonian systems, the
Lagrangian $L(q,\dot q)$ (or the Hamiltonian
$H(q,p)=\frac12\sum_{ij}\kappa^{ij}p_ip_j$ in the cotangent
representation of the flow) is always the first integral of the
system.

Suppose that a Lie group $K$ acts by isometries on $(Q,\kappa)$
(the Lagrangian $L$ is $K$- invariant) and let  $\xi_Q$ be the
vector field on $Q$ associated to the action of one-parameter
subgroup $\exp(t\xi)$, $\xi\in\mathfrak k=T_{\mathrm{Id}} K$. The
following version of the Noether theorem holds (see \cite{AKN,
BKMM}): if $\xi_Q$ is a section of the distribution $\mathcal D$
then
\begin{equation}
\frac{d}{dt}\left( \frac{\partial L}{\partial \dot q},\,\xi_Q
\right) =0. \label{moment_map}
\end{equation}

On the other side, let $\xi_Q$ be transversal to $\mathcal D$, for
all $\xi\in\mathfrak k$. In addition, suppose that $Q$ has a
principal bundle structure $\pi: Q\to Q/K$ and that $\mathcal D$
is the collection of horizontal spaces of a principal connection.
Then the nonholonomic geodesic flow defined by $(Q,\kappa,\mathcal
D)$ is called a $K$-{\it Chaplygin system}. The system
(\ref{Hamilton}) is $K$-invariant and reduces to the tangent
bundle $T(Q/K)=\mathcal D/K$ (for the details see \cite{Koi, BKMM,
CCLM, EKR}).

The equations (\ref{Hamilton}) are not Hamiltonian. However, in
some cases they have a rather strong property -- an invariant
measure (e.g, see \cite{AKN, Koz2, BZ}). Within the class of
$K$-Chaplygin systems, the existence of an invariant measure is
closely related with their reduction to a Hamiltonian form after
an appropriate time rescaling $d\tau=\N dt$ (see \cite{Ch2, St,
FeJo, CCLM, EKR}).

Veselov and Veselova \cite{VeVe1, VeVe2} constructed nonholonomic
systems on unimodular Lie groups with right-invariant
nonintegrable constraints and left-invariant metrics, so called
{\it LR systems}, and showed that they always possess an invariant
measure. Similar integrable nonholonomic problems on Lie groups,
with left and right invariant constraints, are studied in
\cite{FeKo, Jo1, Jo2, Bl_Z, FeJo2}. Recently, a nontrivial example
of a nonholonomic LR system, which can be regarded also as a
generalized Chaplygin system ({\it $n$-dimensional Veselova rigid
body problem} \cite{VeVe1, FeKo}) such that Chaplygin reducibility
theorem is applicable for any dimension is given by Fedorov and
Jovanovi\' c \cite{FeJo}.

It appears that LR systems  can be viewed as a limit case of
certain artificial systems (L+R systems) on the same group, which
also possess an invariant measure (see Fedorov \cite{Fe1}). The
latter systems do not have a straightforward mechanical or
geometric interpretation and arise as a ``distortion'' of a
geodesic flow on $G$ whose kinetic energy is given by a sum of a
left- and right-invariant metrics.

A class of L+R systems on $G$ can be seen as a reduction of a
class of nonholonomic systems defined on the  {\it semi-direct
product} of the group $G$ and a vector space $V$ (see Theorems 3,
4 in Schneider \cite{Sch}). We shall prove that an arbitrary L+R
system on $G$ can be obtained as a reduction of a coupled
noholonomic LR system defined on the {\it direct} product $G\times
G$.

One of the best known examples of integrable nonholonomic systems
with an invariant measure is the celebrated Chaplygin sphere which
describes a dynamically non-symmetric ball rolling without sliding
on a horizontal plane and the center of the mass is assumed to be
at the geometric center \cite{Ch1}. It is interesting that the
Chaplygin's sphere appears within both constructions. In the
construction described in \cite{Sch} one should take for the
configuration space the Lie group of Euclidean motion $SE(3)$,
that is the semi-direct product of $SO(3)$ and $\R^3$ \cite{Sch}.
On the other side, Chaplygin sphere is a LR system on the {\it
direct} product $SO(3)\times \R^3$ (e.g., see \cite{Fe2}). This
was a starting point in considering the coupled nonholonomic LR
systems below.

\paragraph{Outline and results of the paper.} In Section 2 we recall the
definition and basic properties of LR and L+R systems. We define
the coupled LR systems and show that any L+R system can be obtain
as a reduction of an appropriate coupled LR system (Sections 3,
4). An example of a coupled LR system on $G\times \g$ is given,
which provides an alternative generalization of the Chaplygin
sphere problem  (Section 4, system (\ref{GSR-eq}) in Section 6).

In Section 5 we study a $n$-dimensional variant of the spherical
support system introduced by Fedorov \cite{Fe}: the motion of a
dynamically nonsymmetric ball $\cal S$ with the unit radius around
its fixed center that touches $N$ arbitrary dynamically symmetric
balls whose centers are also fixed, and there is no sliding at the
contacts points.

Recall that the rubber rolling of the sphere $S^2$ over some other
fixed convex surface in $\R^3$ means that that in the addition to
the constraint given by the condition that the velocity of the
contact point is equal to zero, we have no-twist condition that
rotations about the normal to the surface are forbidden. The
rubber rolling of the dynamically non-symmetric sphere over the
another sphere, considered as a Chaplygin system on the bundle
$SO(3)\times S^2\to S^2$ (where $SO(3)$ acts diagonally on the
total space), as well as the Hamiltonization in sphero-conical
variables of $S^2$ is given by Koiller and Ehlers \cite{KE}. The
integrable cases are found by Borisov and Mamaev \cite{BM4}. In
particular, when the radius of the fixed sphere tends to infinity,
we get the rubber rolling of the sphere over the plane ({\it
rubber Chaplygin sphere}). The Chaplygin reducing multiplier for
the rubber Chaplygin sphere is given in  \cite{EKR}.

By the analogue, we define the $n$-dimensional rubber spherical
support system with additional no-twist conditions at the contact
points. It appears that both systems fits into the construction of
coupled LR systems. Similarly as for the 3-dimensional spherical
support system studied in \cite{Fe}, we prove that the
3-dimensional rubber spherical support system is integrable
(Section 5).

Finally in Section 6 we consider the $n$-dimensional {rubber
Chaplygin sphere problem} describing the rolling without slipping
and twisting of an $n$-dimensional ball on an $(n-1)$-dimensional
 hyperplane $\mathcal H$ in ${\mathbb
R}^{n}$ as coupled LR systems on the direct product
$SO(n)\times\R^{n-1}$. It appears that the rubber Chaplygin sphere
is a $SO(n-1)\times \R^{n-1}$-Chaplygin system closely related to
the $n$-dimensional nonholonomic Veselova  problem, which allows
as to prove the existence of the Chaplygin multiplier for a
specially chosen inertia operator of the ball. In particular, when
$n=3$, the multiplier exist for any inertia tensor of the ball,
and reduces to the one obtained in \cite{EKR, KE}.

\section{Preliminaries}

\paragraph{LR systems.}
\textit{LR system} on a Lie group $G$ is a nonholonomic geodesic
flow of a left-invariant metric and right-invariant nonintegrable
distribution $\mathcal D\subset TG$ (see \cite{VeVe1, VeVe2}).
Through the paper we suppose that all considered Lie groups $G$
have bi-invariant Riemannian metrics, or equivalently
$\Ad_G$-invariant Euclidean scalar products $\langle \cdot,\cdot
\rangle$ on corresponding Lie algebras. In particular, Lie groups
$G$ are unimodular. 

Let $\g=T_{\mathrm{Id}}G$ be the Lie algebra of $G$. In what
follows we shall identify $\g$ and $\g^*$ by means invariant
scalar product $\langle \cdot,\cdot\rangle$, and $TG$ and $T^*G$
by the bi-invariant metric. For clearness, we shall use the symbol
$\omega$ for the elements in $\g$ and the symbol $m$ for the
elements in $\g^*\cong \g$.

The Lagrangian is defined by
$
L(g,\dot g)=\frac12\langle I\omega,\omega\rangle,
$
where $\omega=g^{-1}\cdot \dot g$ is the {\it angular velocity in the moving frame.}
Here $I: \g\to \g$ is a symmetric positive definite
(with respect to $\langle \cdot,\cdot\rangle$) operator.
The corresponding left-invariant metric will be denoted by $(\cdot,\cdot)_{I}$.
The distribution $\mathcal D$ is determined by its restriction
${\D}$ to the Lie algebra and it is nonintegrable if and only if
$\D$ is not a subalgebra of $\g$. Let $\h$ be the orthogonal
complement of $\D$ with respect to $\langle \cdot,\cdot \rangle$
and let $a_1,\dots,a_{k}$ be a orthonormal base of $\h$. Then the
right-invariant constraints can be written as
$$
\langle\Omega,\mathfrak h\rangle=\langle \omega , \h^g\rangle=0,
\qquad \h^g=\Ad_{g^{-1}}(\h)=g^{-1}\cdot \h \cdot g,
$$
or, equivalently,
\begin{equation}
\langle \alpha_i,\omega \rangle=0, \quad
\alpha_i=\Ad_{g^{-1}}(a_i), \quad i=1,\dots,k. \label{Rconstr}
\end{equation}
Here $\Omega=\Ad_g(\omega)=\dot g \cdot g^{-1}$ represents {\it angular velocity in the space}.

Equations (\ref{Hamilton}) in the left trivialization take the form
\begin{eqnarray}
&&\dot m=[m,\omega]+\sum_{i=1}^{k} \lambda_i \alpha_i, \label{lr1}\\
&&\dot g=g\cdot \omega,  \label{lr2}
\end{eqnarray}
where
$
m={\partial L}/{\partial \omega}=I\omega\in\g^*
$
is the {\it angular momentum in the body frame.}

The Lagrange multipliers $\lambda_i$ can be found by
differentiating the constraints (\ref{Rconstr}). They are actually
defined on the whole phase space $T^*G$ and we can consider the
system (\ref{lr1}), (\ref{lr2}) on $T^*G$ as well (see
\cite{VeVe2}). The constraint functions $\langle \alpha_i,\omega
\rangle$ are then integrals of the extended system and the
nonholonomic geodesic flow is just the restriction of (\ref{lr1}),
(\ref{lr2}) onto the invariant submanifold (\ref{Rconstr}).

Instead of  (\ref{lr1}), (\ref{lr2}),  one can consider the closed
system consisting of (\ref{lr1}) and
\begin{equation}
\dot \alpha_i=[\alpha_i,\omega],\qquad i=1,\dots,k, \label{lr4}
\end{equation}
on the direct product $\g^{1+k}=\{(m,\alpha^1,\dots,\alpha^k)\}$.
Let $ I^{-1}|_{h^g}=\pr_{\h^g} \circ I^{-1} \circ \pr_{\h^g}$,
where $\pr_{\h^g}$ is the orthogonal projection to $\h^g$. Then
the system (\ref{lr1}), (\ref{lr4}) has an invariant measure with
density $\mu=\sqrt{\det (I^{-1}|_{\h^g})}=$
 $\sqrt{\det\left(\langle
I^{-1}(\alpha_i),\alpha_j\rangle\right)}$ (see \cite{VeVe2}).

Also, since for $\xi\in \g$, the associate vector field $\xi_G$ of
the left $G$-action is right invariant and the momentum mapping of
the left action equals to $M=\Ad_g(m)$ ({\it angular momentum in
the space}), the LR system  (\ref{lr1}), (\ref{lr2}) has the
Noether conservation laws:
\begin{equation}
\frac{d}{dt} \langle\Ad_g(m),\,\xi\,\rangle=0, \qquad \xi\in \D.
\label{Noether}\end{equation}

If the linear subspace $\mathfrak{h}$ is the Lie algebra of a
subgroup $H\subset G$, then the Lagrangian $L$  and the
right-invariant distribution  $\mathcal D$  are invariant with
respect to the left $H$-action. As a result, the LR system  can
naturally be regarded as a $H$-Chaplygin system \cite{FeJo}.

\paragraph{Geodesic flow on $G$ with L+R metric.}
In addition to the nondegenerate linear operator ${I}$ defining
the left-invariant metric $(\cdot,\cdot)_{I}$, introduce a
constant symmetric linear operator $ \Pi^{0}:\; \g\rightarrow \g $
defining a  right-invariant metric $(\cdot,\cdot)_{\Pi }$ on the
$n$-dimensional compact Lie group $G$: for any vectors $u, v\in
T_{g} G$ we put $(u,v)_{\Pi }=\langle u g^{-1},\Pi ^{0} v
g^{-1}\rangle$. We take the sum of both metrics and consider the
corresponding geodesic flow on $G$ described by the  Lagrangian
$$
L=\frac12 \langle \omega,{I}\omega \rangle +\frac12 \langle g
\omega g^{-1},\Pi^0 \, g\omega g^{-1} \rangle = \frac12 \langle
\omega,{I}\omega \rangle + \langle\omega,\Pi^g \omega\rangle,
$$
where $\Pi^g={\Ad}_{g^{-1}}\Pi^0 {\Ad}_g$. We can also consider
the case when $\Pi^g$ is not positive definite, but the total
inertia operator ${\cal B}= {I}+\Pi^g$ is nondegenerate and
positive definite on the whole group $G$.

The geodesic motion on the group is described by the Euler--Poincar\'e  equations
\begin{equation}
\dot m= [ m, \omega] + g^{-1} \frac{\partial L}{\partial g},\qquad
m =\frac{\partial L}{\partial\omega}={\cal B}\omega,
\label{i5.3}
\end{equation}
together with the kinematic equation $\dot g=g\cdot\omega$.

In order to find explicit expression for $g^{-1} ({\partial
L}/{\partial g})$, we first note that for any $\xi\in {\mathfrak
g}$, $ \langle \xi,g^{-1} ({\partial L}/{\partial g})\rangle =
v_\xi (L), $ where $v_\xi$ is the left-invariant vector field on
$G$ generated by $\xi$. Since the metric $(\cdot,\cdot)_{{I}}$ is
left-invariant, we have
\begin{gather*}
v_\xi (L) = \frac 12 v_\xi (\langle\omega,\Pi\omega\rangle)=
\frac12 \langle \omega,\Pi{\rm ad\,}_{\xi}\omega +{\rm
ad\,}_{\xi}^T\Pi\omega\rangle=
\langle\Pi\omega,[\xi,\omega]\rangle = \langle \xi, {\rm
ad\,}_\omega\, \Pi\, \omega \rangle.
\end{gather*}
As a result, $g^{-1} ({\partial L}/{\partial g})={\rm ad\,}_\omega\, \Pi\, \omega$.

Also, in view of the definition of $\Pi$, its evolution is given by $n\times n$
matrix equation
\begin{equation}
\dot\Pi=\Pi{\rm ad\,}_\omega+{\rm ad\,}_\omega^T\Pi.
\label{i5.4}
\end{equation}
Since $\langle\cdot,\cdot\rangle$ is $\Ad_G$ invariant scalar product, we have ~${\rm ad\,}_\omega^T=-{\rm
ad\,}_\omega$, and~$\dot\Pi=[\Pi,\ad_\omega]$.

Equations (\ref{i5.3}), (\ref{i5.4}) form a closed system on the
space $\mathfrak g\times {\bf Symm}(n)$ with the coordinates
$\omega_i, \Pi_{ij}$ ($\omega=\sum_i \omega_i e_i$,
$\Pi=\sum_{i\le j} \Pi_{ij} e_i\otimes e_j$), where
$e_1,\dots,e_n$ is a orthonormal base of $\g$.

\paragraph{L+R systems.}
Following Fedorov \cite{Fe1}, consider the  equations (\ref{i5.3})
modified by rejecting the term $g^{-1} ({\partial l}/{\partial
g})$. As a result, we obtain the another system
\begin{equation}
\frac d{dt}({\cal B}\omega)= [{\cal B}\omega,\omega], \qquad \dot
g=g\cdot\omega,  \qquad{\cal B}={I}+\Pi  \label{L+R}
\end{equation}
on $TG$, or the system
\begin{equation}
\frac d{dt}({\cal B}\omega)= [{\cal B}\omega,\omega], \qquad \frac
d{dt} \Pi=\Pi\, {\rm ad\,}_\omega+{\rm ad\,}_\omega^T \, \Pi
\label{i5.8}
\end{equation}
on the space $\mathfrak g\times {\bf Symm}(n)$. This is generally
not a Lagrangian system, and, in contrast to equations
(\ref{i5.3}), (\ref{i5.4}), it possesses the ``momentum'' integral
~$\langle{\cal B}\omega,{\cal B}\omega\rangle$. In view of the
structure of the kinetic energy, we shall refer to the system
(\ref{L+R}) (or (\ref{i5.8})) as {\it L+R system} on $G$
\cite{Fe1}.

The L+R system (\ref{i5.8}) possesses also the kinetic energy
integral $\frac 12\langle\omega,{\cal B}\omega\rangle$ and an
invariant measure (in coordinates $\omega_{i}$, $\Pi_{ij}$)
$\mu\,d\omega_1 \wedge\cdots\wedge \omega_n\land
d\Pi_{11}\wedge\cdots\wedge\Pi_{nn}$
 with density
$\mu=\sqrt{\det ({I}+\Pi )}$ (see \cite{FeRCD, Fe1}).

As mentioned above, a nonholonomic LR system on a Lie group $G$
can be obtained as a limit case of a certain L+R system on this
group. Indeed, suppose that the operator defining a
right-invariant metric on $G$ is {\it degenerate} and has the form
$\Pi=\epsilon({\alpha}_1\otimes{\alpha}_1+\cdots+{\alpha}_{k}
\otimes{\alpha}_{k})$, $k <n$, $\epsilon={\rm const}>0$, where, as
in (\ref{Rconstr}), ${\alpha}_1, \dots, {\alpha}_{k}$ are
orthonormal right-invariant vector fields $\alpha_i=g^{-1}\cdot
a_i \cdot g$, $a_i=\mbox{const}\in \g$. The L+R system
(\ref{i5.8}) on the space $(\omega,{\alpha}_1,\dots,{\alpha}_{k})$
can be represented in form
\begin{equation} \label{resolved}
{I} \dot\omega= {I} ({I}+\Pi)^{-1} [I\omega,\omega],\quad
\dot\Pi=\Pi{\rm ad\,}_\omega+{\rm ad\,}_\omega^T\Pi.
\end{equation}

Then the following statement holds (see \cite{Fe1}). As
$\epsilon\rightarrow \infty $, the equations (\ref{resolved})
transform to the equations  with multipliers (\ref{lr1}) and
constraints (\ref{Rconstr}), where $m={I}\omega$.



\section{Coupled nonholonomic LR Systems}

Define a {\it coupled nonholonomic LR system} on the direct
product $G\times G_1$ ($G=G_1$) as a LR system given by the
Lagrangian function
\begin{equation}
L=\frac12  \langle I\omega,\omega\rangle  + \frac12 D \langle
\mathbf w,\mathbf w \rangle \label{c-lagr}
\end{equation}
and right-invariant constraints
\begin{eqnarray}&& \langle \Omega,\h_0\rangle=0, \label{c-0}\\
 &&\langle \Omega+\rho_i \mathbf W,\h_i\rangle =0, \qquad
i=1,\dots,q, \label{c-constr}
\end{eqnarray}
where $\h_i$, $i=1,\dots,q$ are mutually orthogonal linear
subspaces of $\g$.

Here $(\omega,\mathbf w)=(g^{-1}\dot g,g_{1}^{-1}\dot g_1)$, is
the angular velocity in the body and $(\Omega,\mathbf
W)=\Ad_{(g,g_1)}=(\Ad_{g}(\omega),\Ad_{g_1}(\mathbf w))$ is the
angular velocity in the space, $(g,g_1)\in G\times G_1$. The
constant $D$ is greater than zero, while $\rho_i$, $i=1,\dots,q$
are arbitrary non-zero, real parameters.

The Lagrangian (\ref{c-lagr}) in the second variable is
right-invariant as well. It is convenient to write the equations
of motion both in the left-trivialization (in variables $g$ and
$\omega$) and right-trivialization (in variables $g_1$ and
$\mathbf W$)
\begin{equation}
T(G\times G_1)\approx G\times G_1 \times \g \times
\g_1=\{(g,g_1,\omega,\mathbf W)\}. \label{left-right}
\end{equation}

Then the right-invariant distribution $\mathcal D\subset T(G\times
G_1)$ is given by
$$
\mathcal D=\{(g,g_1,\omega,\mathbf W) \, \vert\, \langle
\Ad_g(\omega),\h_0\rangle=0,\, \,\langle \Ad_g(\omega)+\rho_i
\mathbf W,\h_i\rangle =0,\,\,\,i=1,\dots,q\, \}.
$$

Let $\h^g_i=\Ad_{g^{-1}}(\h_i)=g^{-1}\cdot \h_i\cdot g$ and let
$\pr_{\h^g_i}:\g \to \h^g_i$  be the orthogonal projections,
$i=0,\dots,q$.

\begin{prop}\label{ct1}
The admissible path $(g(t),g_1(t),\omega(t),\mathbf W(t))$ is a
motion of the nonholonomic LR system (\ref{c-lagr}), (\ref{c-0}),
(\ref{c-constr}) if it satisfies equations
\begin{eqnarray}
&\mathcal B\dot\omega & =\,\,\,[I\omega,\omega]-(\mathcal B^{-1}|_{h^g_0})^{-1}\pr_{\h^g_0}\mathcal B^{-1}([I\omega,\omega]),\label{I} \\
& D\dot{\mathbf W}&=\,\,-\sum_{i=1}^q \frac{D}{\rho_i}\pr_{\h_i}(\Ad_g \,\dot\omega),\label{II} \\
& \dot g &=\,\,\, g\cdot \omega,\label{III}\\
& \dot g_1& =\,\,\, \mathbf W \cdot g_1.\label{IIII}
\end{eqnarray}
where $ \mathcal B=I+ \Pi=I+\sum_{i=1}^q
{D}/{\rho_i^2}\pr_{\h^{g}_i}$ and $ \mathcal
B^{-1}|_{h^g_0}=\pr_{\h_0^g}\circ \mathcal B^{-1}
\circ\pr_{\h_0^g} \, : \h^g_0 \rightarrow \h^g_0. $
\end{prop}

\noindent{\it Proof.} The equations of a motion in the
right-trivialization (or in the space frame) read
\begin{eqnarray}
 &\dot M &= \quad\sum_{i=0}^q \Lambda_i \label{Ia}\\
&D\dot{\mathbf W}&=\quad \sum_{i=1}^q \rho_i\Lambda_i,\label{IIa}\\
&\dot g&=\quad\Omega\cdot g, \\
&\dot g_1&=\quad\mathbf W \cdot g_1,
\end{eqnarray}
where the Lagrange multipliers (reaction forces) $\Lambda_i$
belong to $\mathfrak h_i$ ($i=0,1,\dots,q$) and $M=\Ad_g(I\omega)$
is the first component of {\it angular momentum in the space
frame} (the second component is $M_1=D\mathbf W$).

Differentiating the constraints (\ref{c-constr}), from (\ref{IIa}) we obtain
$$
\frac{d}{dt}\langle \Omega+\rho_i \mathbf W,\h_i\rangle= \langle
\dot \Omega+\rho_i \dot{\mathbf W},\h_i\rangle=\langle
\dot\Omega+\rho_i\sum_{j=1}^q \frac{\rho_j}{D}
\Lambda_j,\h_i\rangle=\langle \dot\Omega+ \frac{\rho^2_i}{D}
\Lambda_i,\h_i\rangle=0,
$$
that is
\begin{equation}
\Lambda_i=-\frac{D}{\rho^2_i}\pr_{\h_i}(\dot\Omega), \qquad
i=1,\dots,q. \label{c-multiplier}
\end{equation}
The equation (\ref{II}) follows from (\ref{IIa}),
(\ref{c-multiplier}) and the relation
\begin{equation}
\dot\Omega=\Ad_g\,\dot\omega.
\label{IDENTITY}
\end{equation}

From (\ref{c-multiplier}) and identities (\ref{IDENTITY}),
$\pr_{\h^g_i}=\Ad_{g^{-1}}\pr_{\h_i}\Ad_g$ and
$$
\Ad_{g^{-1}}\dot M=I\dot\omega+[\omega,I\omega],
$$
the equation
(\ref{Ia}) in the left-trivialization takes the form
\begin{equation}
\mathcal B\dot\omega =[I\omega,\omega]+\lambda_0, \qquad
\lambda_0=\Ad_{g^{-1}}(\Lambda_0). \label{Ic}
\end{equation}

Now it remains to find the Lagrange multiplier $\lambda_0$.
Differentiating (\ref{c-0}) we get
$$
 \langle \dot \Omega, \h
\rangle=\langle \Ad_{g}(\dot \omega),\h\rangle=\langle
\dot\omega,\h_0^g\rangle=0.
$$ Whence, according (\ref{Ic}) it
follows $ \lambda_0=-(\mathcal
B^{-1}|_{h^g_0})^{-1}\pr_{\h^g_0}\mathcal
B^{-1}([I\omega,\omega])$. The proof is complete. \hfill $\Box$

\medskip

The Lagrangian (\ref{c-lagr}) as well as constraints
(\ref{c-constr}) are right ($\{\mathrm{Id}\} \times
G_1$)-invariant and the equations (\ref{I}), (\ref{II}),
(\ref{III}) can be seen as a reduction of the system to
$$
\bar{\mathcal D}=\mathcal D/(\{\mathrm{Id}\}\times
G_1)=\{(g,\omega,\mathbf W)\, \vert\, \langle
\Ad_g(\omega),\h_0\rangle=0,\,\, \langle \Ad_g(\omega)+\rho_i
\mathbf W,\h_i\rangle =0,\,\, i=1,\dots,q\, \}\,.
$$

Let $\mathcal D_0\subset TG$ be the right-invariant distribution
defined by (\ref{c-0}).

\begin{thm}\label{ct2}
The equations (\ref{I}), (\ref{II}), (\ref{III}) on $\bar{\mathcal
D}$ are reducing to the following system on $\mathcal D_0\subset
TG$:
\begin{equation}
\frac{d}{dt}(\mathcal B\omega)=[\mathcal B\omega,\omega]-(\mathcal
B^{-1}|_{h^g_0})^{-1}\pr_{\h^g_0}\mathcal
B^{-1}([I\omega,\omega]), \qquad \dot g=g\cdot \omega.
\label{Ib}
\end{equation}
\end{thm}

\noindent{\it Proof.} The equations (\ref{I}) and (\ref{III}) form
a closed system on $\mathcal D_0$.  If $(g(t),\omega(t))$ is a
solution of (\ref{I}), (\ref{III}), then one can easily
reconstruct the motion of $\mathbf W$. Let
\begin{equation}
\mathfrak k =(\h_1+\dots+\h_q)^\perp. \label{KK}
\end{equation}
From (\ref{III}) we have
\begin{equation}\frac{d}{dt} \pr_{\mathfrak k} \mathbf W=0,  \label{c-low}
\end{equation}
while the $\h_i$-components of the angular velocity $\mathbf W$
are determined from the constraints (\ref{c-constr}):
$$
\pr_{\h_i} \mathbf W=-{1}/{\rho_i}\pr_{\h_i} \Ad_g(\omega), \qquad
i=1,\dots,q.
$$

Now, let $a_1,\dots,a_{k_j}$ be the orthonormal base of $\h_j$.
Then
$\alpha_1=\Ad_{g^{-1}}(a_1),\dots,\alpha_{k_j}=\Ad_{g^{-1}}(a_{k_j})$
will be the orthonormal base of $\h^g_j$. We have
$$
\pr_{\h^g_j}(\omega)=\sum_{i=1}^{k_j} \alpha_i \otimes \alpha_i \,
\omega=\sum_{i=1}^{k_j} \langle \alpha_i,\omega\rangle \alpha_i.
$$

Whence, by using (\ref{lr4}) and the identity
$\langle \omega,[\alpha_i,\omega]\rangle=0$, we obtain
\begin{eqnarray*}
\frac{d}{dt}\Big(\pr_{\h^g_j}(\omega)\Big)&=&\sum_{i=1}^{k_j}
\Big( \langle \dot \omega, \alpha_i \rangle \alpha_i + \langle
\omega, [\alpha_i,\omega] \rangle \alpha_i+
\langle  \omega, \alpha_i \rangle [\alpha_i,\omega]\Big) \nonumber\\
&=& \pr_{\h^g_j}(\dot \omega)+\sum_{i=1}^{k_j} \langle \omega,
\alpha_i \rangle [\alpha_i,\omega]. \label{der}
\end{eqnarray*}
The above equation implies that (\ref{I}), (\ref{III}) can be
rewritten in the form  (\ref{Ib}).  \hfill $\Box$

\

The derivation of $\langle \mathcal B\omega,\omega \rangle$ along
the flow is: $\frac{d}{dt}\langle \mathcal
B\omega,\omega\rangle=2\langle [\mathcal B,\omega],\omega \rangle
+ 2\langle \lambda_0,\omega \rangle. $ The first term is equal to
zero since $\langle \cdot,\cdot\rangle$ is a $\Ad_G$-invariant
scalar product, while the second term is equal to zero from the
constraint (\ref{c-0}). We can refer to $ L_{red}=\frac12\langle
\mathcal B\omega,\omega\rangle $ as to the {\it reduced
Lagrangian}, or {\it reduced kinetic energy}. If $\pr_{\mathfrak
k}\mathbf W \equiv 0$, the reduced kinetic energy coincides with
the kinetic energy of the reconstructed motion on the whole phase
space.

From the equation (\ref{Ia}) we also get the linear conservation
law
\begin{equation}
\frac{d}{dt} \left(\pr_{\mathfrak k_0} \Ad_g(I\omega) \right)=0,
\qquad {\rm where} \qquad \mathfrak
k_0=(\h_0+\h_1+\dots+\h_q)^\perp \,. \label{c-low*}
\end{equation}

The integrals (\ref{c-low}) and (\ref{c-low*}) are actually
Noether integrals (\ref{Noether}) of the system. The other
Noethers integrals are trivial:
$$
\frac{d}{dt}\left(\pr_{\h_i}
\Ad_g(I\omega)-\frac{D}{\rho_i}\pr_{\h_i}\mathbf W\right)=0,
\qquad i=1,\dots,q.
$$

\begin{remark}\label{REMARK}{\rm
If $\h_0=0$, i.e., we do not impose the constraint (\ref{c-0}),
the reduced system is an L+R system on the Lie group $G$
\begin{equation}
\frac{d}{dt}(\mathcal B\omega)=[\mathcal B\omega,\omega], \qquad
\dot g=g\cdot \omega. \label{Ib*}
\end{equation}
Further suppose
that (\ref{KK}) is the Lie algebra of the closed Lie subgroup
$K\subset G$ and that linear subspaces $\h_i$ are
$\Ad_K$-invariant:
$$
 \Ad_k \h_i=\h_i, \qquad k\in K, \qquad i=1,\dots, q.
$$
Then, since $\h^{kg}_i=\h^g_i$, $k\in K$, the L+R equations
(\ref{Ib*}) are {\it left} $K$-invariant and we can reduce them to
$Q\times \g$, where $Q=G/K$ is the homogeneous space, with respect
to the left-action of $K$.}\end{remark}

\begin{remark}\label{CR}{\rm
In the case when $\h_0$ is the Lie algebra of a closed subgroup
$H\subset G$, $ \h_1+\h_2+\dots+\h_q=\g$ and linear spaces $\h_i$
are $\Ad_H$ invariant, then the coupled LR system (\ref{c-lagr}),
(\ref{c-0}), (\ref{c-constr}) is  $(H\times G_1)$-Chaplygin system
with respect to the action:
$$
(a,b) \cdot (g,g_1)=(ag,g_1b^{-1}), \quad (a,b)\in H\times G_1\,.
$$
The reduced space $\mathcal D/(H\times G_1)$ is the tangent bundle
of the homogeneous space $G/H$.
 }\end{remark}

\begin{thm}
An arbitrary L+R system (\ref{L+R}) can be seen as a reduction of
an appropriate coupled LR system.
\end{thm}

\noindent{\it Proof.} Let $e_1,\dots,e_n$ be the orthonormal base
of $\mathfrak g$ in which the symmetric operator $\Pi^0$ has the
diagonal form: $ \Pi^0=\sum_{i=1}^n \sigma _i \, e_i\otimes e_i. $
Then the right invariant term in (\ref{L+R}) reads $
\Pi=\Pi^g=\sum_{i=1}^n \sigma _i \, e_i^g\otimes e_i^g$, where
$e_i^g$ are given by
\begin{equation}
e_1^g=\Ad_{g^{-1}}(e_1),\dots,e_n^g=\Ad_{g^{-1}}(e_n).
\label{moving-base}
\end{equation}

Consider the $2$-coupled nonholonomic LR system (\ref{c-lagr}),
(\ref{c-constr}), where $q=n$ and $\h_i$ are the lines in the
directions of $e_i$, $i=1,\dots,n$. We can choose parameters $D$,
$\rho_i$, such that $\sigma_i=D/\rho_i^2$, $i=1,\dots,n$. The
system represents a $\{\mathrm{Id}\}\times G_1$- Chaplygin system
with reduced equations of the required form (\ref{L+R}). \hfill
$\Box$

\section{$N$-Coupled Systems}

There is a straightforward generalization of the construction to
the case when we have coupling with $N$ different Lie groups, that
is the configuration space is the direct product $G\times G_1
\times \dots \times G_N$ and the Lagrangian is
\begin{equation}
L=\frac12 \langle I \omega,\omega \rangle + \frac12\sum_{i=1}^N
D_i \langle \mathbf w_i,\mathbf w_i \rangle_i, \label{c-N-lagr}
\end{equation}
where $\langle \cdot,\cdot\rangle_i$ are $\Ad_{G_i}$ invariant
scalar products on Lie algebras $\g_i=T_{\mathrm Id} G_i$,
$i=1,\dots,N$.

Let us fix a base $e_1,\dots,e_n$ of $\g$ and some bases
$f_1,\dots,f_{d_i}$ of $\g_i$ ($d_i=\dim\g_i$). Let
$$
A_i: \g \to \R^{p_i}, \quad B_i: \g_i \to \R^{p_i}, \quad
i=1,\dots,N.
$$
be the linear mappings with matrixes $[A_i]$ ($p_i\times n$) and
$[B_i]$ ($p_i\times d_i$) in the above bases. In addition, we
suppose that the ($p_i\times p_i$)-matrixes
$$
[C_i]=[B_i][B_i]^T, \quad i=1,\dots,N
$$
are invertible. Consider the right invariant constraints given by
\begin{equation}
A_i\Omega+B_i\mathbf W_i=0, \qquad i=1,\dots,N. \label{c-N-constr}
\end{equation}
Here, $\omega$, $\mathbf w_i$ and $\Omega$, $\mathbf W_i$ are
velocities in the left and right trivializations, respectively and
$D_i > 0$, $i=1,\dots,N$ are real parameters.

Let $[\Omega]$, $[\mathbf W_i]$ denote the column matrix,
representing $\Omega$ and $\mathbf W_i$ in the chosen bases. We
have $[\omega]_g=[\Omega]$, where $[\xi]_g$ is the column,
representing $\xi\in\g$ in the base (\ref{moving-base}).

In the right-trivialization, the equation in $\mathbf W_i$ reads
\begin{equation}
D_i [\dot{\mathbf W}_i]=[B_i]^T[\lambda_i], \label{c-N-right}
\end{equation}
where $[\lambda_i]$ is the Lagrange multiplier ($p_i\times
1$)-matrix. Differentiating the constraints (\ref{c-N-constr}),
from (\ref{c-N-right}) we get
\begin{equation}
[\lambda_i]=-D[C_i]^{-1}[A_i][\dot \Omega], \qquad i=1,\dots,N.
\label{c-N-multiplier}
\end{equation}

Repeating the arguments of Theorems \ref{ct1} and \ref{ct2}, the
considered $N$-coupled nonholonomic system reduces to the L+R
system
$$
\frac{d}{dt}(\mathcal B\omega)=[\mathcal B\omega,\omega], \qquad
\dot g=g\cdot \omega,
$$
where $\mathcal B\omega =I\omega + \Pi\omega$, and $\Pi\omega$ in the matrix form, relative to the base
(\ref{moving-base}), is given by
$$
[\Pi\omega]_g = \sum_{i=1}^N D_i [A_i]^T[C_i]^{-1}[A_i][\omega]_g.
$$

As above, one can easily incorporate an additional right invariant
constraint of the form (\ref{c-0}).

\paragraph{LR systems on $G\times\g\times \cdots\times \g$.}
As an example, consider the case where $G_i$ are all equal to the
Lie algebra $\g$ considered as a Abelian group, $\langle
\cdot,\cdot \rangle_i=\langle \cdot,\cdot \rangle$ and the
constraints (\ref{c-N-constr}) are given by
\begin{equation}
[\Omega,\Gamma_i]+\rho_i\mathbf W_i=0,\qquad i=1,\dots,N,
\label{GC-c}
\end{equation}
where $\Gamma_i$ are fixed elements of the Lie algebra $\g$ and
$\rho_i$ are real parameters. Note that, since $G_i=\g$ is Abelian
group, the angular velocities coincide with the usual velocity:
$\dot \xi_i=\mathbf W_i=\mathbf w_i$, $\xi\in \g$.

The equations of a motion in the right-trivialization read
\begin{eqnarray}
 &&\dot M =\sum_{i=0}^N [\Lambda_i,\Gamma_i], \qquad \dot g=\Omega\cdot g \label{GC1}\\
&&  D_i\dot{\mathbf W_i}=\rho_i\Lambda_i,\qquad \quad \dot
\xi_i=\mathbf W_i, \label{GC2}
\end{eqnarray}
where $M=\Ad_g(I\omega)$. This is a $\{\mathrm{Id}\}\times
\g^N$--Chaplygin system and it is reducible to $TG$.
Differentiating the constraints (\ref{GC-c}), from (\ref{GC2}) we
get the Lagrange multipliers
\begin{equation*}
\Lambda_i=-\frac{D}{\rho^2_i}[\Gamma_i,\dot\Omega], \qquad
i=1,\dots,N. \label{GS-m}
\end{equation*}

Therefore, the equations (\ref{GC2}) in the left-trivialization
take the form
\begin{equation*}
I\dot\omega =[I\omega,\omega]-\sum_{i=1}^N
\frac{D_i}{\rho_i^2}[[\gamma_i,\dot\omega],\gamma_i], \qquad \dot
g=g\cdot \omega. \label{GC3}
\end{equation*}
where $\gamma_i=\Ad_{g^{-1}}(\Gamma_i)$, $i=1,\dots, N$. Next,
from the identities
$$
\frac{d}{dt}[[\gamma_i,\omega],\gamma_i]=[[\gamma_i,\dot
\omega],\gamma_i]-[[[\gamma_i,\omega],\gamma_i],\omega], \qquad
i=1,\dots,N,
$$
we obtain the following proposition

\begin{prop}
The reduced equations of the $N$--coupled nonholonomic system
(\ref{c-N-lagr}), (\ref{GC-c}) are given by the L+R system
\begin{equation}
\frac{d}{dt}(\mathcal B\omega)=[\mathcal B\omega,\omega], \qquad
\dot g=g\cdot\omega, \label{GC-eq}
\end{equation}
where
$$
 \mathcal B\omega =I\omega+\sum_{i=1}^N
\frac{D_i}{\rho_i^2}[[\gamma_i,\omega],\gamma_i].
$$
\end{prop}

\begin{remark}{\rm
Nonholonomic systems on semi-direct products $G\times_\sigma V$,
where $\sigma$ is a representation of the Lie group $G$ on the
vector space $V$ are studied in  Schneider \cite{Sch}. Proposition
4.1 can be derived from Theorem 3 given in
\cite{Sch}.}\end{remark}

\section{Spherical Support}

Consider the motion of a dynamically nonsymmetric ball $\cal S$ in
$\R^n$ with the unit radius around its fixed center. Suppose that
the ball touches $N$ arbitrary dynamically symmetric balls whose
centers are also fixed, and there is no sliding at the contacts
points. We call this mechanical construction {\it the spherical
support}. For $n=3$ spherical support is defined by Fedorov
\cite{Fe, Fe1}.

The configuration space is $SO(n)^{N+1}$: the matrixes $g, g_i\in
SO(n)$ map the frames attached to the ball $\mathcal S$ and the
$i$th $i$th peripheral ball to the fixed frame, respectively. The
Lagrangian is of the form (\ref{c-N-lagr}), where for
$\langle\cdot,\cdot\rangle$ we take the scalar product
proportional to the Killing form
\begin{equation}
 \langle X,Y\rangle=-\frac12\tr(XY),
\label{KF}
\end{equation} the angular velocities $\omega$,
$\Omega$, $\mathbf w_i$, $\mathbf W_i$ of the balls are defined as
above, $I: so(n) \to so(n)$ is the inertia tensor of the ball
$\cal S$ and $D_{i},\rho_{i}\in {\mathbb R}$ are the central
inertia moment and the radius of the $i$th peripheral ball.

Let $\Gamma_{i}\in \R^n$ be the unit vector {\it fixed in the
space} and directed from the center $C$ of the ball $\cal S$ to
the point of contact with the $i$th ball. Nonholonomic constraints
express the absence of sliding at the contact points. This means
that  velocity of the point of contact of the ball $\cal S$ with
the $i$th ball, in the space frame, is the same as the  velocity
of the corresponding point on the $i$th ball.

\begin{figure}[h,t]
\begin{center}
\includegraphics[width=.7\textwidth]{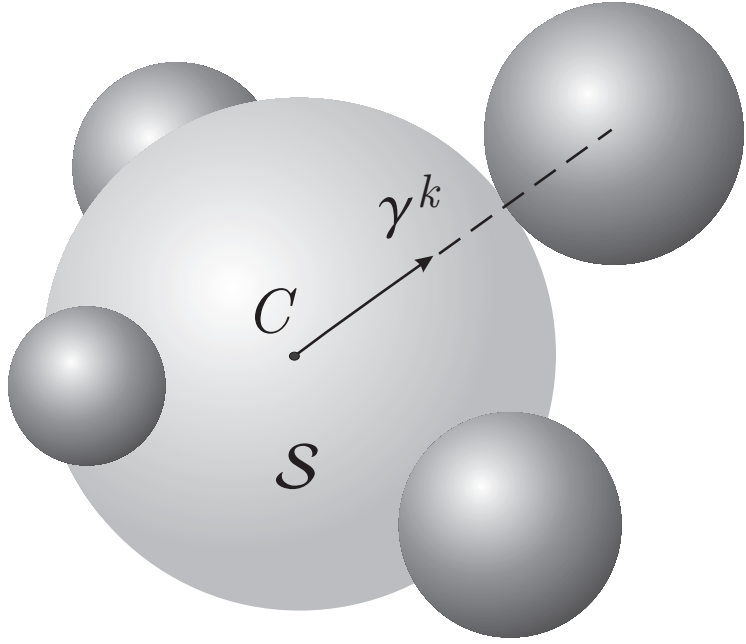}
\caption{\footnotesize The spherical support} \label{st.fig}
\end{center}
\end{figure}

Consider the fixed point on the ball $\mathcal S$ with coordinates
$r$ and $R$ in the body and space frames, respectively. Then the
velocity of the point $r$ in space is given by the Poisson
equation (e.g, see \cite{FeKo}) $V=\dot R=\frac{d}{dt}(g\cdot
r)=\dot g \cdot g^{-1}\cdot  g\cdot r=\Omega R.$ Therefore, the
velocity of the contact point with the $i$th peripheral ball is
given by $\Omega\Gamma_i$. Similarly, the velocity of the
corresponding contact point of the $i$th ball in the space frame
is given by $-\rho_i\mathbf W\Gamma_i$ and the constraints are
\begin{equation}
\Omega \Gamma_i+\rho_i\mathbf W_i\Gamma_i=0, \qquad i=1,\dots,N.
\label{ss-constr}\end{equation}

We see that the $n$-dimensional spherical support is actually a
$N$-coupled LR system studied in the previous section. Let
\begin{equation}
\gamma_i=g^{-1}\Gamma_i, \qquad i=1,\dots N
\label{gamma}
\end{equation}
be the contact points of $\mathcal S$ with the $i$th ball
($i=1,\dots,N$) in the frame attached to the ball $\mathcal S$.
Then the right-invariant constraints (\ref{ss-constr}) can be
rewritten in the form
\begin{equation}
\langle \Omega +\rho_i\mathbf W_i,\h_i\rangle =0, \qquad
i=1,\dots,N, \label{ss-constr*}
\end{equation}
where
$$
\h_i=\R^n\wedge \Gamma_i, \qquad
\h_i^g=\Ad_{g^{-1}}(\h_i)=\R^n\wedge \gamma_i, \qquad i=1,\dots,N
$$
are linear (no mutually orthogonal) subspaces of the Lie algebra
$so(n)$.

From the identity $ \pr_{\h_i^{g}}\dot\omega=(\dot
\omega\gamma_i)\wedge \gamma_i=\dot \omega\, \gamma_i \otimes
\gamma_i+\gamma_i \otimes \gamma_i \, \dot\omega, $ the equations
of the motion become
\begin{eqnarray*}
&&I\dot\omega = [I\omega,\omega]-\sum_{i=1}^N \frac{D_i}{\rho_i^2}
\left(\dot \omega \, \gamma_i \otimes \gamma_i+\gamma_i \otimes \gamma_i \,\dot\omega\right),\qquad \; \dot g = g\cdot \omega\\
&& D_i\dot{\mathbf W}_i =-\frac{D_i}{\rho_i} \left(\dot\Omega\,
\Gamma_i \otimes \Gamma_i+\Gamma_i \otimes \Gamma_i\,
\dot\Omega\right), \qquad\qquad \dot g_i = \mathbf W_i \cdot
g_i,\quad  i=1,\dots,N.
\end{eqnarray*}

We have the conservation laws
$$
\dot{\mathbf W}_i-\dot{\mathbf W}_i\, \Gamma_i \otimes \Gamma_i -\Gamma_i \otimes \Gamma_i\, \dot{\mathbf W}_i=0,\qquad i=1,\dots,N,
$$
which together with the right $(\{\mathrm{Id}\}\times
SO(n)^N)$-symmetry lead to the following statement

\begin{prop}
The spherical support system reduces to the L+R flow
\begin{equation}
\frac{d}{dt}(\mathcal B\omega)=[\mathcal B\omega,\omega], \qquad
\dot g=g\cdot\omega, \label{ss-red}
\end{equation}
where $\mathcal B\omega =I\omega+\sum_{i=1}^N {D_i}/{\rho_i^2}
\left(\omega\, \gamma_i \otimes \gamma_i+\gamma_i\otimes \gamma_i
\,\omega\right)$ and $\gamma_i$ are defined by  (\ref{gamma}).
\end{prop}

One can say that the reduced system (\ref{ss-red}) on $TSO(n)$
describes the free rotation of a ``generalized
Euler top'', whose tensor of inertia is a sum of two components: one is fixed in
the body and the other one is fixed in the space.

Note that the vectors $\gamma_i$ in the frame attached to the ball
$\mathcal S$ satisfy the Poisson equations (e.g., see \cite{FeKo})
\begin{equation}
\dot\gamma_i=-\omega \gamma_i, \qquad i=1,\dots,N.
\label{ss-poisson}
\end{equation}
By introducing $\mathcal X_i=\gamma_i \otimes \gamma_i$, from
(\ref{ss-poisson}) we obtain
\begin{equation}
\dot {\mathcal X}_i=[\mathcal X_i,\omega], \qquad i=1,\dots,N.
\label{ss-poisson*}
\end{equation}

Combining (\ref{ss-red}) and (\ref{ss-poisson*}) we get family of
integrals - the coefficients of the polynomials
\begin{equation}
\tr(\mathcal B\omega + \sum_{i=1}^N \mu^i {\mathcal X}_i)^k,
\qquad k=1,\dots,n. \label{ss-integrals}\end{equation}

For $n=3$ the system is integrable by the Euler--Jacobi theorem,
and its generic invariant manifolds are two-dimensional tori (see
\cite{Fe, Fe1}).

\begin{remark}{\rm
If the positions of peripheral balls are mutually orthogonal
$$
(\Gamma_i,\Gamma_j)=(\gamma_i,\gamma_j)=\delta_{ij}, \qquad 1\le i,j \le N \le n,
$$
then the components of $\gamma_i$ can be seen as redundant
coordinates on the Stiefel variety $V(n,N)=SO(n)/SO(n-N)$. The
system is invariant with respect to the $SO(n-N)$ action,
representing the rotations in the space orthogonal to
$\Span\{\gamma_1,\dots,\gamma_N\}$. The $SO(n-N)$-reduced system
on $TSO(n)/SO(n-N) \cong V(n,N)\times so(n)$ is given by Poisson
equations (\ref{ss-poisson}) and the first equation in
(\ref{ss-red}). }\end{remark}

\paragraph{Rubber spherical support.}
Now consider the {\it rubber spherical support system} in $\R^n$.
The analogue of rubber rolling  is that, in addition to the
constraints (\ref{ss-constr*}), the rotations of the ball
$\mathcal S$ and $i$th peripheral ball around the vector
$\Gamma_i$ are the same: \begin{equation} \langle \Omega-\mathbf
W_i, \mathfrak k_i \rangle=0, \qquad i=1,\dots,N,
\label{rr-constr}
\end{equation}
where
$$ \mathfrak k_i=\h_i^\perp, \qquad \mathfrak k_i\cong so(n-1).
$$

Since $\pr_{\mathfrak k_i}=\mathbf I-\pr_{\h_i}$ we get

\begin{prop}
The rubber spherical support system is described by the equations
\begin{eqnarray} &&\frac{d}{dt}(\mathcal B^*\omega) = [\mathcal
B^*\omega,\omega],\label{rr-red}\\
&& D_i\dot{\mathbf W}_i =D_i\dot\Omega-D_i\frac{1+\rho_i}{\rho_i}
\left(\dot\Omega\, \Gamma_i \otimes \Gamma_i+\Gamma_i \otimes
\Gamma_i\, \dot\Omega\right),\quad i=1,\dots, N,\label{rr-W}\\
&&\dot g=g\cdot \omega, \label{rr-g}\\
 &&\dot g_i =
\mathbf W_i \cdot g_i,\quad  i=1,\dots,N,
\end{eqnarray}
where
$$
\mathcal B^* \omega =I\omega+(D_1+\dots+D_N)\omega+\sum_{i=1}^N
D_i\frac{1-\rho_i^2}{\rho_i^2} \left(\omega\, \gamma_i \otimes
\gamma_i+\gamma_i\otimes \gamma_i \,\omega\right).
$$
\end{prop}

The equations (\ref{rr-W}) are trivial since $\mathbf W$ can be
expressed in terms of $\Omega$ from constraints (\ref{ss-constr*})
and  (\ref{rr-constr}).

As above, we get family of geometric integrals that can be
expressed as the coefficients of the polynomials
\begin{equation}
\tr(\mathcal B^*\omega + \sum_{i=1}^N \mu^i \mathcal X_i)^k,
\qquad k=1,\dots,n. \label{rr-integrals}\end{equation}

For $n=3$, among the reduced kinetic energy $\frac12\langle
\mathcal B^*\omega,\omega\rangle$ and integrals
(\ref{rr-integrals}) there are four independent one.

\begin{thm}
For $n=3$, the rubber spherical support system (\ref{rr-red}),
(\ref{rr-g}) is solvable by the Euler--Jacobi theorem and its
generic invariant manifolds are two-dimensional tori.
\end{thm}

\section{Rubber Chaplygin Sphere}

Following \cite{FeKo,Fe2},  consider the generalized Chaplygin
sphere problem of an $n$-dime\-nsional ball of radius $\rho$,
rolling without slipping on an $(n-1)$-dimensional hyperspace
$\mathcal H$ in ${\mathbb R}^{n}$. For the configuration space we
take the {\it direct product} of Lie groups $SO(n)$ and $\R^n$,
where $g\in SO(n)$ is the rotation matrix of the sphere (mapping
frame attached to the body to the space frame) and  $r\in {\mathbb
R}^n$ is the position vector of its center $C$ (in the space
frame). For a trajectory $(g(t),r(t))$ define angular velocities
$$
\omega=g^{-1}\dot g, \qquad \Omega=\dot g g^{-1}, \quad \mathbf
w=\mathbf W=\dot r. $$
The Lagrangian of the system is then given
by
\begin{equation}
L=\frac12\langle I\omega,\omega\rangle+\frac12 m(\mathbf w,\mathbf w).
\label{ch-lagr}
\end{equation}
Here  $I: so(n) \to so(n)$ and $m$ are the inertia tensor and mass
of the ball, $\langle\cdot,\cdot\rangle$ is given by (\ref{KF})
and $(\cdot,\cdot)$ is the Euclidean scalar product.

Let $\Gamma\in{\mathbb R}^{n}$ be a {\it vertical } unit vector
(considered in the fixed frame) orthogonal to the hyperplane
$\mathcal H$ and directed from $\mathcal H$ to the center $C$. The
condition for the sphere to role without slipping leads that the
velocity of the contact point is equal to zero:
\begin{equation} \label{ch-constr}
-\rho\mathbf{\Omega}\Gamma+\mathbf W =0 \, .
\end{equation}

This is a right-invariant  nonholonomic constraint of the form
(\ref{c-N-constr}). If we take the fixed orthonormal base $
E_1=(1,0,\dots,0,0)^T, \dots$, $E_n=(0,0,\dots,0,1)^T, $ such that
$\Gamma=E_n$, then the constraint (\ref{ch-constr}) takes the form
$$
\dot r_i=\rho\Omega_{in}, \quad i=1,\dots,n-1, \quad \dot r_n=0,
\quad {\rm where}\quad \Omega_{ij}=\langle \Omega, E_i \wedge E_j\rangle.
$$

The last constraint is holonomic, and for the physical motion we
take $r_n=\rho$. From now on we take $SO(n)\times \R^{n-1}$ for
the configuration space of the rolling sphere, where $\R^{n-1}$ is
identified with the affine hyperplane $\rho\Gamma+\mathcal H$.

Let $\h\subset so(n)$ be the linear subspace $\h=\R^n\wedge\Gamma$
and $\mathfrak k\cong so(n-1)$ its orthogonal complement in
$so(n)$. Define the {\it rubber Chaplygin sphere} as a Chaplygin
sphere (\ref{ch-lagr}), (\ref{ch-constr}) subjected to the
additional right-invariant constraints
\begin{equation}
\langle \Omega, \mathfrak k \rangle=\langle \omega, \mathfrak k^g
\rangle =0, \quad \mathfrak k^g=\Ad_{g^{-1}}\mathfrak k, \quad
\Longleftrightarrow\quad \Omega_{ij}=0, \quad 1\le i < j \le n-1 ,
\label{ch-rubber}
\end{equation}
describing the no-twist condition at the contact point. As a
result, the distribution
$$
\mathcal D=\{(g,r,\omega,\mathbf W) \, \vert\, \langle
\omega,\mathfrak k^g\rangle=0,\, \mathbf W=\rho
\Ad_g(\omega)\Gamma \}.
$$
is right $SO(n)\times \R^{n-1}$ as well as the left $SO(n-1)\times
\R^{n-1}$ invariant ($SO(n-1)$ is the subgroup of $SO(n)$ with the
Lie algebra $\mathfrak k$). Moreover, the rubber Chaplygin sphere
is a ($SO(n-1)\times \R^{n-1}$)-Chaplygin system.

Let $\gamma$ be the vertical vector in the frame attached to the
ball $\gamma=g^{-1}\Gamma$. Then
$$
\mathfrak h^g=\Ad_{g^{-1}}(\h)=\R^n \wedge \gamma=:\mathfrak
h^\gamma, \qquad \mathfrak k^g=\Ad_{g^{-1}}(\mathfrak
k)=(\R^n\wedge\gamma)^\perp=:\mathfrak k^\gamma
$$
and the reduced space $\mathcal D/(SO(n-1)\times\R^{n-1})$ is the
tangent bundle $TS^{n-1}$ of the sphere which can be identified by
the position of $\gamma$.

The equations in the right-trivialization read
\begin{eqnarray}
&&\dot M=\Lambda_0-\rho \Lambda_1 \wedge \Gamma, \qquad \dot g=\Omega\cdot g, \label{ch-right-eq}\\
&&m\dot{\mathbf W}=\Lambda_1, \qquad \qquad \,\,\,\quad \dot
r=\mathbf W, \label{ch-r-w}
\end{eqnarray}
where $M=\Ad_g(I\omega)$ is the ball angular momentum in the space
and  $\Lambda_0\in\mathfrak h$, $\Lambda_1\in\R^n$ are Lagrange
multipliers. From (\ref{ch-constr}) and (\ref{ch-r-w}) we find $
\Lambda_1= m\rho \dot\Omega\Gamma. $ On the other hand
\begin{equation}
\Lambda_1 \wedge \Gamma=m\rho(\dot\Omega\Gamma)\wedge
\Gamma=m\rho\left( \dot\Omega\, \Gamma \otimes \Gamma + \Gamma
\otimes \Gamma \,\dot\Omega\right)=m\rho\pr_\h(\dot\Omega).
\label{ch}\end{equation}

Whence, we can write equations (\ref{ch-right-eq}) as a closed
system on $\mathcal D_0\subset TSO(n)$, where $\mathcal D_0$ is
the right-invariant distribution defined by (\ref{ch-rubber})
(reduction of $\R^{n-1}$-symmetry). From (\ref{IDENTITY}),
(\ref{ch}) and the relation $ \pr_{\h^\gamma}(\dot\omega)=(\dot
\omega \cdot \gamma)\wedge \gamma={\dot \omega \,}
\gamma\otimes\gamma+ \gamma \otimes \gamma{\,\dot\omega} $, in the
left-trivialization of $TSO(n)$ the reduced system takes the form
$$
I\dot\omega  =[I\omega,\omega]-m\rho^{2}( {\dot \omega \,}
\gamma\otimes\gamma+ \gamma \otimes \gamma{\,\dot\omega}
)+\lambda_0, \quad \dot g=g\cdot \omega,
$$
where $\lambda_0=\Ad_{g^{-1}}(\Lambda_0)$.
Let
\begin{equation} \label{K}
\mathbf k=I\omega+m\rho^2\pr_{\mathfrak h^\gamma}
\omega=I\omega+m\rho^2( {\omega \,} \gamma\otimes\gamma+ \gamma
\otimes \gamma{\,\omega} )\in so(n)^*
\end{equation}
be the angular momentum of the ball relative to the contact point
(see \cite{FeKo}). Then we have:

\begin{prop}
The motion of the rubber Chaplygin sphere, in variables $\omega,
g$, is described by
\begin{equation}
\dot{\mathbf k}=[\mathbf k,\omega]+\lambda_0, \qquad \dot
g=g\cdot\omega,\label{ch-r-red}
\end{equation}
or, in variables $\omega,\gamma$, by equations
\begin{equation}
\dot{\mathbf k}=[\mathbf k,\omega]+\lambda_0, \qquad
\dot\gamma=-\omega\gamma. \label{ch-r-red2}
\end{equation}
The Lagrange multiplier matrix $\lambda_0$ belongs to $\mathfrak
k^\gamma$ and is determined from the constraint (\ref{ch-rubber}).
\end{prop}

\paragraph{Reduction and Hamiltonization.}
From the constraints (\ref{ch-rubber}), the momentum (\ref{K})
equals to $\mathbf k=I\omega +m\rho^2 \omega$. Therefore, as in
the 3--dimensional case \cite{EKR, BM3}, the equations
(\ref{ch-r-red}) are equivalent to the motion of a rigid body
about the fixed point with the left-invariant kinetic energy given
by the inertia operator $I+m\rho^2\mathbf I$ and constraint
(\ref{ch-rubber}) ({$n$-dimensional Veselova rigid body problem}
\cite{VeVe1, FeKo}).

Now we simply follow \cite{FeJo}. The reduced Lagrange-d'Alambert
equations of the rubber Chaplygin sphere (\ref{ch-lagr}),
(\ref{ch-constr}), (\ref{ch-rubber}) on
$$
TS^{n-1}\cong  \mathcal D/(SO(n-1)\times \R^{n-1})\cong \mathcal D_0/SO(n-1)
$$ are given by
\begin{equation}
\left( \frac{\partial L_{red}}{\partial \gamma} - \frac{d}{dt}
\frac{\partial L_{red}}{\partial \dot \gamma}\, , \xi\right) =
\langle (I+m\rho^2)\Phi(\gamma,\dot \gamma), (\mathbf
I-\pr_{\h^\gamma}) [\Phi(\gamma,\dot \gamma),\Phi(\gamma, \xi)]
\rangle, \label{reduced}
\end{equation}
for all virtual displacements $\xi\in T_\gamma S^{n-1}$ (see
\cite{FeJo}). Here $ \Phi(\gamma,\dot \gamma)=\gamma \wedge \dot
\gamma$ is the momentum mapping of the right $SO(n)$-action on the
round sphere $S^{n-1}$ and the reduced Lagrangian  is given by
\begin{equation}
L_{red}(\gamma,\dot \gamma)=\frac12 \langle
(I+m\rho^2)\Phi(\gamma,\dot\gamma),\Phi(\gamma,\dot\gamma)\rangle.
 \label{reduced_lagrangian}
 \end{equation}

After the Legendre transformation
$$
p=\frac{\partial L_{red}}{\partial \dot \gamma}=\frac{\partial
L_{reg}}{\partial \Phi}\,\frac{\partial \Phi}{\partial
\dot\gamma}=m\rho^2\dot\gamma-I\Phi \cdot \gamma
$$
we can also write the reduced Lagrange-d'Alambert equations as a
first-order system on the cotangent bundle $T^*S^{n-1}$ which is
realized as a subvariety of ${\mathbb R}^{2n}=(q,p)$ defined by
constraints $(\gamma,\gamma)=1$, $(\gamma,p)=0$ (since $I\Phi$ is
skew-symmetric, the momentum $p$ satisfies $(\gamma,p)=0$). The
system takes the symmetric form
\begin{equation}
\dot \gamma= -\Phi(\gamma,\dot\gamma(\gamma,p)) \gamma, \qquad
\dot p=-\Phi(\gamma,\dot\gamma(\gamma,p)) p,
\label{legendre}\end{equation} where
$\dot\gamma=\dot\gamma(\gamma,p)$ is the inverse of the Legendre
transformation.

Let $\sigma$ be the canonical volume $2(n-1)$-form on
$T^*S^{n-1}$. Then we have (see \cite{FeJo}):

\begin{prop}
\label{redLR} The reduced system (\ref{legendre}) on $T^*S^{n-1}$
possesses an invariant measure
$$
1/\sqrt{ \det (I+m\rho^2\mathbf I |_{\h^\gamma}) }\,\, \sigma,
\qquad I+m\rho^2 \mathbf I |_{\h^\gamma}=\pr_{\h^\gamma}\circ
(I+m\rho^2\mathbf I) \circ \pr_{\h^\gamma}\, .
$$
\end{prop}

Furthermore, as it follows from \cite{FeJo}, with the operator $I$
defined on the bi-vectors $X\wedge Y$ by a diagonal matrix
$A=\diag(A_1,\dots,A_n)$ by
\begin{equation}
I(X\wedge Y)=AX \wedge AY- m\rho^2 \, X\wedge Y,
\label{inertia_tensor}
\end{equation}
the Chaplygin reducibility is applicable for any dimension:

\begin{thm} \label{main}
\begin{description}
\item{(i)} If the inertia operator is given by
(\ref{inertia_tensor}), the density of an invariant measure in
Proposition \ref{redLR} takes the following simple form
$$
(A\gamma,\gamma)^{-(n-2)/2}.
$$

\item{(ii)} Under the time substitution $d\tau
=1/\sqrt{(A\gamma,\gamma) }\, dt$ the reduced system
(\ref{reduced}) (or (\ref{legendre})) becomes a Hamiltonian system
describing a geodesic flow on $S^{n-1}$ with the Lagrangian
\begin{equation} \label{L^*}
L^{\ast }\left(\gamma,\frac{d \gamma}{d\tau}\right)=\frac12\left[
\bigg(A \frac{d\gamma}{d\tau},
\frac{d\gamma}{d\tau}\bigg)(A\gamma,\gamma) - \bigg(A\gamma,
\frac{d\gamma}{d\tau}\bigg)^2 \right]\, .
\end{equation}

\item{(iii)} For $A$ with distinct eigenvalues, the latter system
is algebraic completely integrable and  generic invariant
manifolds are $(n-1)$-dimensional tori.

\item{(iv)} Moreover, the $SO(n-1)$-reconstruction of the motion
is solvable: the generic trajectories of the system
(\ref{ch-r-red}) are straight-lines (but not uniform) over
$(n-1)$-dimensional invariant tori.
\end{description}
\end{thm}

The complete integration is presented in \cite{FeJo}. Given a
solution $(g(t),\omega(t))$ of the system (\ref{ch-r-red}), the
reconstruction of $r$-variable simply follows from the integration
of the constraint (\ref{ch-constr})
\begin{equation*}
r(t)-r(t_0)=\rho\int_{t_0}^t \Ad_{g(t)}\omega(t)\Gamma \, dt \,.
\end{equation*}

In the case $n=3$, under the isomorphism between $so(3)$ and
${\mathbb R}^3$
\begin{equation}
{\mathbf{\omega} }_{ij}=\varepsilon_{ijl} \omega_l , \quad  {\bf
{k}}_{ij}=\varepsilon_{ijl} \mathbf k_l, \label{iso}\end{equation}
from (\ref{ch-r-red2}) we obtain the classical rubber Chaplygin's
ball equations \cite{EKR}
\begin{equation}
\dot{\vec{\mathbf k}}=\vec{\mathbf k}\times\vec
\omega+\lambda\vec\gamma, \qquad  \dot{\vec \gamma}=\vec
\gamma\times\vec\omega, \label{Chap}\end{equation} where $\lambda$
is determined from the constraint $(\vec\omega,\vec\gamma)=0$ and
$\vec{\mathbf k}=I \vec\omega+ m\rho^2\vec \omega-m\rho^2 (\vec
\omega,\vec\gamma)\vec \gamma=I\vec\omega+m\rho^2\vec\omega$.

For $n=3$, the relation (\ref{inertia_tensor}) defines a generic
inertia tensor. Thus the rubber Chaplygin sphere in $\R^3$ is
integrable. Indeed, let $I:\R^3 \to \R^3$ be an arbitrary inertia
tensor. Under the isomorphism (\ref{iso}), the matrix $A$ is
determined from (\ref{inertia_tensor}) via:
$$
A=\Delta (I+m\rho^2\mathbf I)^{-1}, \qquad
\Delta=\sqrt{\det(I+m\rho^2\mathbf I)}.
$$

The Hamiltonizaton of the reduced system on $T^*S^2$ is obtained
in \cite{EKR, KE}. The Chaplygin multiplier given in Theorem
\ref{main}
$$
d\tau=dt/\sqrt{\Delta((I+m\rho^2\mathbf I)^{-1}\gamma,\gamma)}
$$
up to the multiplication by a constant, coincides with the
expression obtained in \cite{EKR, KE}.

\paragraph{Remarks on the Chaplygin sphere.}
\begin{description}
\item{$\bullet$} Note that the Chaplygin sphere equations
$$
\dot{\mathbf k}=[\mathbf k,\omega], \qquad
\dot\gamma=-\omega\gamma, \qquad \mathbf k=I\omega+m\rho^2(
{\omega \,} \gamma\otimes\gamma+ \gamma \otimes \gamma{\,\omega} )
$$
coincide with the reduced equations of the spherical support
system for $N=1$, where instead of $D_1/\rho_1$ we should put
$m\rho^2$. This is not the case for rubber analogues of the
systems.

\item{$\bullet$} Borisov and Mamaev \cite{BM, BM3} proved that the
classical Chaplygin rolling sphere problem is Hamiltonian after an
appropriate time rescaling. Recently, the Hamiltonization of the
homogeneous Chaplygin rolling sphere problem in $\R^n$ is given in
\cite{HN}, while the Hamiltonization of the non-homogeneous
reduced Chaplygin sphere problem is obtained in \cite{Jo5}.

\item{$\bullet$} Let us turn back to the coupled LR system
described in Proposition 4.1. Take $N=1$ and denote
$\Gamma_1=\Gamma$, $\gamma_1=\gamma$, $D_1=m$, $\rho_1=1/\rho$.
The system (\ref{GC-eq}) is additionally $G_\Gamma$--invariant,
where $G_{\Gamma}\subset G$ is the isotropy group of $\Gamma$. Let
$\mathcal O=G/G_{\Gamma}$ be the adjoint orbit of $\Gamma$. Then
(\ref{GC-eq}) reduces to $\mathcal O \times \g$:
\begin{equation}
\dot{\mathbf k}=[\mathbf k,\omega], \qquad \dot \gamma
=[\gamma,\omega], \qquad \mathbf k=\mathcal
B\omega=I\omega+m\rho^2 [[\gamma,\omega],\gamma].
 \label{GSR-eq}
\end{equation}

For $G=SO(3)$ we reobtain the equations of a motion of the
Chaplygin sphere in $\R^3$. Thus the system (\ref{GSR-eq}) can be
seen as an alternative generalization of the Chaplygin sphere
problem.
\end{description}

\subsection*{Acknowledgments}
I am greatly thankful to Yuri N. Fedorov for useful discussions.
Also, he created Figure 4.1. The research was supported by the
Serbian Ministry of Science, Project 144014, Geometry and Topology
of Manifolds and Integrable Dynamical Systems.

\

Bo\v zidar Jovanovi\' c

Mathematical Institute SANU

Serbian Academy of Science and Art

Kneza Mihaila 36, 11000, Belgrade, Serbia

e-mail: bozaj@mi.sanu.ac.rs

\end{document}